\documentclass[a4paper,12pt]{article}
\usepackage{epsfig}
\usepackage{amssymb}
\begin{document}
\thispagestyle{empty}

\newcommand{\etal}  {{\it{et al.}}}  
\def\Journal#1#2#3#4{{#1} {\bf #2}, #3 (#4)}
\def\PRD{Phys.\ Rev.\ D}
\def\NIMA{Nucl.\ Instrum.\ Methods A}
\def\PRL{Phys.\ Rev.\ Lett.\ }
\def\PLB{Phys.\ Lett.\ B}
\def\EPJ{Eur.\ Phys.\ J}
\def\IEEETNS{IEEE Trans.\ Nucl.\ Sci.\ }
\def\CPCD{Comput.\ Phys.\ Commun.\ }


\bigskip

{\Large\bf
\begin{center}
Lesson from a soluble model of critical point and primary photons 
\end{center}
}
\vspace{0.1 cm}

\begin{center}
{ G.A. Kozlov  }
\end{center}
\begin{center}
\noindent
 { Bogolyubov Laboratory of Theoretical Physics\\
 Joint Institute for Nuclear Research,\\
 Joliot Curie st., 6, Dubna, Moscow region, 141980 Russia  }
\end{center}
\vspace{0.1 cm}

 \begin{abstract}
 \noindent
 {The critical point in particle physics at high temperature is studied through the ideal gas of scalars, the dilatons, in the model that implies the spontaneous breaking of an approximate scale symmetry. We consider the dynamical system of identical particles weakly interacting to each other. The critical point with the temperature as a function of a dilaton mass is established. The fluctuation of particle density grows up very sharply at critical point.
Our results also suggest that the critical point may be identified through the fluctuation  in yield of primary photons induced by conformal anomaly of strong and electromagnetic sectors of matter.}


\end {abstract}




\bigskip

{\bf 1. Inroduction}

The spontaneous breaking of symmetry has a major role in particle physics and cosmology where the phase transitions (PT) can occur at extreme conditions (e.g., high enough temperature, baryonic density, chemical baryon potential). The critical point (CP) of PT corresponds to an initial thermal state (of matter) which is invariant under the conformal group $G$. This state has a nontrivial couplings to other states which are the physical matter ones. The latter not to possess the invariance under $G$, but they are invariant under the chiral symmetry group $H$ triggered by $G$.

In the early Universe the light massive and/or massless states are emerged. The critical phenomena if occurred are considered here through quantum PT with  Bose-Einstein condensation (BEC) of the scalar  field (the dilaton) in an ideal Bose-gas. In this case, the condensation takes place in a single zero mode that suggests the breaking of conformal symmetry. 
All scales in particle physics and cosmology  are subjects of dilatons.
The properties of quarks and gluons accompanying by the scalar condensate at high both temperature and baryon chemical potential are the keys for understanding the evolution of the Universe.
There are  also more or less the same pattern for experiments with heavy ion collisions at early times. One can imply the existence of Goldstone-like modes: the scale symmetry is broken explicitly resulting with an appearance of a scalar, the dilaton, in the spectrum, accompanying by $\pi$-mesons as a consequence of chiral symmetry breaking. 

It is well-known that quantum effects, e.g. gluon fluctuations, break conformal (scale) invariance. It is seen through the anomaly in the trace of energy-momentum tensor $\theta ^{\mu}_{\mu} = \partial_{\mu} S^{\mu} \neq 0$, where the dilatation current $S^{\mu} = \theta ^{\mu\nu} x_{\nu}$ does not conserved itself with respect to the scale transformations of coordinates $x_{\mu}\rightarrow \omega x_{\mu}$ ($\omega$ is an arbitrary constant). If the explicit breaking of scale symmetry does not play the dominant role, the spontaneous breaking of chiral symmetry may imply the spontaneous breaking of the approximate scale symmetry. The dilaton appeared as a pseudo-Goldstone boson is associated with the chiral condensate occurred in the region where the gauge coupling constants are slowly running and an effective fermion coupling constant has reached the critical value [1].  

Because of presence of strong gluon fields, the QCD vacuum is disordered and scale invariance is destroyed by the appearance of  the dimensional scale $M$  
$$M = M_{UV}\,\exp \left [-8\,\pi^{2}/(b_{0}\,g^{2})\right ] $$
with $M_{UV}$ being the ultra-violet (UV) scale, $g$ is the bare gauge coupling constant and $b_{0}$ is the first coefficient in the QCD $\beta$-function, $\beta (M)\neq 0$. 
The breaking of conformal invariance assumes that all the processes are governed by the conformal anomaly (CA) resulting from running coupling constant $g(M)$. 
The derivative of $S_{\mu}$ is proportional to  $\beta$-function and the quark masses. 
For $SU(N_{c})$ gauge theory with $N_{c}$ number of colours and $N_{f}$ number of flavours in the fundamental representation, the $\beta$-function is
$$\beta(\alpha_{s}) \equiv M\frac{\partial\alpha_{s}(M)}{\partial M} \equiv -\frac{b_{0}}{2\pi}\alpha^{2}_{s} - \frac{b_{1}}{(2\pi)^{2}}\alpha^{3}_{s} + ...,$$
where $\alpha_{s} \equiv \alpha_{s} (M)$ is the renormalised gauge coupling constant defined at the scale $M$;  $b_{1}$ is the coefficient. There can be an approximate scale (dilatation) symmetry if $\beta(\alpha_{s})$ is small enough and $\alpha_{s}(M)$ is slowly running with $M$. Theory becomes conformal in the infra-red (IR)  with the non-trivial solution $\alpha_{s}^{\star} = -2\pi\,b_{0}/b_{1}$ (IR fixed point (IRFP) or the Banks-Zaks [2] conformal point) in the perturbative domain if $b_{0} = (11 N_{c}- 2 N_{f})/3$ is small. The latter is happened when $(N_{c}/N_{f}) = 2/11$. Once $N_{f}$ decreases near PT (the value of $\alpha_{s}^{\star}$ increases) the CP is characterised by $\alpha^{c}_{s} < \alpha^{\star}_{s}$ at which the spontaneous breaking of the chiral symmetry is occurred, and the confinement does appear. In the neighborhood of the IRFP the $\beta$-function is approximated by 
$\beta(\alpha_{s}) \simeq -\Delta\alpha_{s} (M/\Lambda)^{\delta} $,
where $0 < \Delta\alpha_{s} = \delta (\alpha_{s}^{\star} - \alpha_{s}^{c}) << \alpha_{s}^{c}$, $\delta \leq O(1)$ [3], $\Lambda$ is the confinement scale. 

The dilatons are unstable, they decay into two photons. It means  CA acts as a source of primary (direct) photons (or soft photons in heavy-ion collisions [4]) not produced in decays of light hadrons. The operator associated with primary photon defines the highest weight of representation of conformal symmetry and this operator obeys the unitarity condition $d\geq j_{1}+j_{2} +2 - \delta_{j_{1}j_{2},0}$ for scaling dimension $d$, where $j_{1}$ and $j_{2}$ are the operator Lorentz spins (primary means not a derivative of another operator).

There is a well-known importance of  primary photons emission for the study of evolution pattern of heavy ion collisions, especially of the very early phase. For example, the correlations of direct photons can shed the light to the space-time distribution of the hot matter prior to freeze-out  [5]. However, the investigation of direct photons is faced with considerable difficulties compared with hadron physics mainly due to small yield of photons emitted directly from the hot dense zone in comparison to the huge background of photons produced by the standard decay of final hadrons (e.g., neutral $\pi$ - mesons). 
If CP is situated in some regions accessible to heavy ion collisions experiments one should identified it through some observables to be discovered experimentally. 
The signature of CP is non-monotonous behaviour of observable fluctuation where the latter increases very crucially.

In this paper, we suggest the novel approach to an approximate scale symmetry breaking with the challenge phenomenology where the primary photons induced by CA  are in fluctuating regime
 The rate of the fluctuation length plays an important role of an indicator of CP achievement if this length grows to become very large. 

{\bf 2. Dilaton field}

We start with the dilaton-pion toy model given by the Lagrangian density (LD)
\begin{equation}
\label{e1}
L = \frac{1}{2} f_{\sigma}^{2} \left  (\partial_{\mu} e^{\sigma}\right )^{2} + \frac{1}{2} f_{\pi}^{2}\, e^{2\sigma}(\partial_{\mu}\bar\pi)^{2} + ...
\end{equation}
which is scale invariant under transformations $x_{\mu}\rightarrow \omega x_{\mu}$  if $\bar\pi (x) = {f_{\pi}}^{-1}\pi (x)$ transforms as $\pi (x) \rightarrow \pi( x e^{\omega})$ and $\sigma$ transforms non-linearly 
\begin{equation}
\label{e2}
\sigma (x) \rightarrow \sigma( x e^{\omega}) + \omega.
\end{equation}
LD (\ref{e1}) has the more suitable form 
\begin{equation}
\label{e3}
L = \frac{1}{2}\left (\partial\chi\right )^{2} + \left (\frac{f_{\pi}}{f_{\chi}}\right )^{2}\,\chi^{2}\,(\partial\bar\pi)^{2} + ...,
\end{equation}
if one make redefinition of the dilaton field 
$\sigma (x) \rightarrow \chi (x) = f_{\chi}\, e^{\sigma (x)}$, which transforms non-linearly under (\ref{e2}). In LD (\ref{e3}), $f_{\chi} = \langle \chi\rangle $ is the order parameter for scale symmetry breaking determined by dynamics of the underlying strong sector. The dimensional parameters $f_{\pi}$ and $f_{\chi}$  enter in the form of c-number ratio $(f_{\pi}/f_{\chi}){^2}$.
The dilatation current $S_{\mu}$ acting on the vacuum  $\vert 0\rangle$ defines $\chi$: $\langle 0 \vert S^{\mu}\vert \chi (p)\rangle = i\,p^{\mu}\,f_{\chi} $, 
$\partial_{\mu} \langle 0\vert S^{\mu} (x)\vert \chi (p)\rangle = \langle 0\vert \theta^{\mu}_{\mu} (x)\vert \chi (p)\rangle = -f_{\chi}\,m_{\chi}^{2}\, e^{-i\,p\,x}, $
where for on-shell case
$\langle 0\vert \theta^{\mu\nu} (x)\vert \chi (p)\rangle = f_{\chi} (p^{\mu} p^{\nu} - g^{\mu\nu} p^{2} ) e^{-i\,p\,x}, \,\,\, p^{2} = m_{\chi}^{2} $.
Here, $m_{\chi}$  is the mass of the dilaton, $p_{\mu}$ is the momentum conjugate to $x_{\mu}$.

Actually, the dilaton is naturally seen in field theory given by LD in the general form (see, e.g., [6])
\begin{equation}
\label{e4}
L = \sum_{i} g_{i}(\zeta)\,O_{i}(x),
\end{equation}
where the local operator $O_{i}(x)$ has the scaling dimension $d_{i}$. LD (\ref{e4}) is scale invariant under dilatation transformations $x^{\mu}\rightarrow e^{\omega} x^{\mu}$ with $O_{i} (x)\rightarrow e^{\omega d_{i}} O_{i}(e^{\omega} x)$ if $g_{i}(\zeta)$ is replaced by $g_{i}(\zeta)\rightarrow g_{i} [ (\zeta\,e^{\sigma})]\,e^{\sigma(4-d_{i})}$, where $\sigma(x)$ as the conformal compensator introduces a flat direction.
Obviously, the theory is invariant under scaling transformation if $d_{i} =4$.
We suppose that at the scale $\geq\Lambda$  the dilaton is formed as the bound state of two gluons, the glueball $\chi = O^{++}$, with the mass $m_{\chi} \sim O(\Lambda)$. 
The $\chi$ representing the gluon composite $\langle A_{\mu\nu}A^{\mu\nu}\rangle$ responsible for the QCD trace anomaly [7] 
 may be understood as the string ring solution  so that the ends of the string meet each other to form a circle with some finite radius.
The characteristic feature of the CP is very sharp increasing of the correlation length $\xi$. The latter describes the fluctuations of the order parameter field $\chi$ near CP, and acts as a regulator in the IR with  $\xi = m_{\chi}^{-1}$. The correlation length $\xi$ is not measured directly, however, it influences the fluctuations of observed particles, e.g., the primary photons to which the critical mode couples. 
Actually, in the vicinity of CP, $\xi$ is much larger than that of a size of the particle interacting region at early times.
The dilaton is a mediator between conformal sector and the Standard Model (SM). However, at high enough temperatures  the role of the mediator can be lost by dilaton because of absence of a conformal anomaly ($\theta ^{\mu}_{\mu} = 0$).

{\bf 3. Dilatons. Quantum statistical states}

We consider the field system containing dilatons as almost ideal weakly interacting gas (e.g., the glueball gas) at finite temperature. 
In case of statistical equilibrium at  temperature $T=\beta^{-1}$ the partition function (statistical sum) for $N$ quantum states (particles) is 
\begin{equation}
\label{e5}
 Z_{N} = Sp\, e^{-H\,\beta}, 
\end{equation}
where $H$ is the Hamiltonian 
$$H = \sum_{1\leq j\leq N} H(j)$$
and $\beta$ in (\ref{e5}) differs from those of the QCD $\beta$ - function.
For the system of dilaton functions $\chi_{f}(x)$ which are regular functions in $f$ representation, one has the equation $H(j)\,\chi_{f}(x_{j}) = F(f)\,\chi_{f}(x_{j})$, where 
\begin{equation}
\label{e6}
H = \sum_{f}\,F(f)\,b^{+}_{f}\,b_{f} = \sum_{f}\,F(f)\,n_{f}
\end{equation}
in terms of  operators of creation $b^{+}_{f}$ and annihilation $b_{f}$; $n_{f}$ is an occupation  number. Here, $F(f) = E(f) - \mu\,Q (f)$ with $E(f)$ being the energy, $\mu$ is the chemical potential, $Q(f)$ is the conserved charge with an average density
$$\langle q\rangle = \frac{1}{\Omega}\,\langle Q\rangle = \frac{1}{\Omega}\,\frac{1}{\beta}\,\frac{\partial}{\partial\mu}\,\ln Z_{N}, $$
where $\Omega$ is the volume of the system. In quantum statistical mechanics where the open system has a thermal contact and a particle interaction with a reservoir, $Q$ is the operator $N_{f}$ of particles of the type $f$ with the mean value $Tr \{\rho\,N_{f}\} = \hat n_{f}\,\Omega$, where $\rho$ is the statistical operator,  $\hat n_{f}$ is the density of particles of the type $f$. Interactions between the dilatons should lead to thermal equilibrium, and in case of large $n_{f}$ - to the formation of BEC.
In principle, operators $b_{f}$ in (\ref{e6}) can be distorted by random quantum fluctuations (e.g., by gluons) through the operator $r_{f}$, $b_{f}\rightarrow b_{f} = a_{f} + r_{f}$, where $a_{f}$ is the bare (annihilation) operator.  The function $Z_{N}$ (\ref{e5}) has the form [8] 
\begin{equation}
\label{e7}
Z_{N} = \sum_{... n_{f} ...,\,\,\sum_{f} n_{f} = N} e^{-\beta\,\sum_{f} F({f})\,n_{f}}.
\end{equation}
Since all the  operators $... n_{f} ..$ commute to each other, they may be clarified through the observables. 
The calculation of (\ref{e7}) meets difficulties because of the condition $\sum_{f} n_{f} = N$ for fixed $N$ and within the limit $N\rightarrow\infty$ in the final stage calculations. The latter condition is important because of particle decays: the glueballs are unstable, hence they  may decay into two primary photons  which are registered as a signal that the vicinity of CP is approached. 

The phase transitions are characterised through the singularities (discontinuities) in the dependence of various observables on the parameters, e.g., temperature, chemical potential etc. 
CP manifests itself through the critical chemical potential $\mu_{c}$ and the critical temperature $T_{c}$. Let us consider the following power series  
\begin{equation}
\label{e8}
P(\bar\mu) = \sum_{N =1}^{\infty} Z_{N}\,\bar\mu^{N},
\end{equation}
which is the scan function on $\mu$ and $\beta$,  where $\bar\mu = \mu/\mu_{c}$.
Having in mind (\ref{e7}) one has
\begin{equation}
\label{e9}
P(\bar\mu) = \sum_{... n_{f} ...} e^{-\beta\sum_{f} F(f)n_{f}}\,\bar\mu^{\sum_{f} n_{f}} = \prod_{f}\left [\sum_{0\leq n < \infty} e^{-\beta F(f)n}{\bar\mu}^{n}\right ] = \prod_{f}\frac{1}{1-\bar\mu e^{-F(f)\beta}}.
\end{equation}
Let us consider for simplicity that $F(f) \geq 0$ in (\ref{e9}).
Actually,  the convergence radius $R$ of the series (\ref{e8}) will not be less than 1. In the vicinity of CP ($\bar\mu \simeq 1$) one has $\mu_{c} < E(f)/Q(f)$ for $E = {\vert\vec p\vert}^{2}/(2\,m_{\chi})$, if the dilaton mass 
$m_{\chi}\rightarrow 0$;  $\vert\vec p\vert$ is the momentum of $\chi$.

Let us consider (\ref{e8})  in the form
\begin{equation}
\label{e10}
\frac{P(\bar\mu)}{\bar\mu^{N}} = \sum_{N^{\prime} = 0}^{\infty} \frac{Z_{N^{\prime}}\,\bar\mu^{N^{\prime}}}{\bar\mu^{N}}
\end{equation}
on the real axis $0 < \bar\mu <R$. Because of positive $Z_{N^{\prime}}$, the function (\ref{e10}) has the only one minimum on $(0,R)$ 
$$\frac{d^{2}}{d\bar\mu^{2}}\left [P(\bar\mu)\,\bar\mu^{-N}\right ] = \sum_{N^{\prime} = 0}^{\infty} (N^{\prime} - N)\,(N^{\prime} - N -1)\,Z_{N^{\prime}}\,\bar\mu^{N^{\prime} - N -2} >0. $$
The function (\ref{e10}) tends to infinity when $\bar\mu\rightarrow 0$ and when $\bar\mu\rightarrow R$. In the interval $(0,R)$ there is a point $\bar\mu = \bar\mu_{0}$ at which  (\ref{e10}) has a single minimum, i.e. 
\begin{equation}
\label{e12}
  \frac{d}{d\bar\mu}\left [P(\bar\mu)\,\bar\mu^{-N}\right ]_{\vert_{\bar\mu =\bar\mu_{0}}} = \sum_{N^{\prime} = 0}^{\infty} Z_{N^{\prime}}\, (N^{\prime} - N)\,\bar\mu^{N^{\prime} - N -1}_{\vert_{\bar\mu = \bar\mu_{0}}} = 0. 
\end{equation}
If one goes alone the vertical axis, (\ref{e10}) has a maximum at $\bar\mu_{0}$. As long as $\bar\mu < \bar\mu_{0}$, no state with $Q \neq 0$ can compete with the vacuum state ($ E = 0$, $Q = 0$) for the role of the ground state.  In case of nonzero baryon density when $\bar\mu > \bar\mu_{0}$, the point $\bar\mu = \bar\mu_{0}$ is the ground state at given $\mu$. 

In the phase space the spectrum of "quasi-momenta" $f$ is almost continuous, and there will be an exact continuous spectrum in the limit $\Omega\rightarrow\infty$. The number $\Delta N$ of different $\Delta f$ in volume $\Omega$ is $(\Delta N/\Delta f) = const\cdot \Omega$. Having in mind that (see (\ref{e9}))
$$ P(\bar\mu) = \exp\left\{ -\sum_{f}\ln \left [ 1-\bar\mu\,e^{-F(f)\,\beta}\right ]\right\}, $$
one can find the asymptotic equality
$$ \sum_{f}\ln \left [ 1-\bar\mu\,e^{-F(f)\,\beta}\right ] = N\,\Phi (\bar\mu),$$
where
$$\Phi (\bar\mu) = const \cdot v\,\beta\,K_{\chi}(\bar\mu), \,\,\, v = \frac{\Omega}{N},$$
$K_{\chi}(\bar\mu)$ is the thermochemical potential of the dilaton $\chi$
\begin{equation}
\label{e14}
K_{\chi}(\bar\mu) = \beta^{-1}\,\int \ln \left [ 1-\bar\mu\,e^{-F(f)\,\beta}\right ] df,
\end{equation}
which gives the contribution to thermodynamic potential 
\begin{equation}
\label{e15}
 K = K_{\chi} + V_{\chi} + \lambda \left ( \frac{f_{\chi}}{2}\right )^{4}.
\end{equation}
Here, $V_{\chi}$ is the potential term in LD of the dilaton 
$$L_{\chi} = \frac{1}{2}\partial_{\mu}\chi\partial^{\mu}\chi - V_{\chi},$$
$$V_{\chi} = \frac{\lambda}{4}\chi^{4}\,\left ( \ln\frac{\chi}{f_{\chi}} -\frac{1}{4}\right ), \,\,\, m_{\chi}^{2} = \frac{d^{2} V(\langle\chi\rangle)}{d\chi^{2}} >  0.$$
The term $\lambda (f_{\chi}/2)^{4}$ in (\ref{e15}) is added so that $K = 0$ at $T = 0$ and $\chi = \langle\chi\rangle = f_{\chi}$. In the vicinity of CP, the free gluons are disfavored as appropriate degrees of freedom in the phase with confinement of quarks. The loops containing the heavy quarks with masses $m_{h}$ can generate radiative corrections to $m_{\chi}^{2}$ of magnitude $\delta m_{\chi}^{2} \sim m_{h}^{2}\,M_{UV}^{2}/(4\,\pi\,f_{\chi})^{2}$. The latter do not influence at CP ($m_{h}\rightarrow 0)$.

The thermodynamic potential (\ref{e15}) does account for dilaton (glueball) and gluon degrees of freedom:
$$K = \theta (\beta_{c} - \beta)\,K_{\chi} (\bar\mu) + \theta (\beta - \beta_{c})\,K_{g} ,$$
where $K_{g}$ is an effective gluon thermodynamic potential with the energy $E_{g} = \sqrt {{\vert \vec p\vert}^2 + m^{2}_{g}}$, $m_{g}$ is an effective gluon mass. $K_{g}$ is model-dependent function and we propose the following its form 
\begin{equation}
\label{e16}
 K_{g} = \beta^{-1}\,\int \ln \left [ 1- e^{-E_{g}(f)\,\beta}\right ] df.
\end{equation}
Both forms (\ref{e14}) and (\ref{e16}) match  each other at CP.

To scan the phase diagram relevant to CP  we use the function
\begin{equation}
\label{e17}
 P(\bar\mu)\,\bar\mu^{-N} = \left [\bar\mu^{-1}\,e^{-\Phi (\bar\mu)}\right ]^{N}.
\end{equation}
CP is a well-defined singularity on the phase diagram $(T, \mu)$. One has to calculate (\ref{e17}) and find the singularity corresponding to the end of the (first order) transition line.
Let us consider the circle $C$ with the radius $r=\bar\mu_{0}$ with the origin at zero. One has
\begin{equation}
\label{e18}
 Z_{N} = \frac{1}{2\,\pi\,i}\int_{C} \frac{P(\bar\mu)}{\bar\mu^{N+1}}\,d\bar\mu \rightarrow \frac{1}{2\,\pi}\int_{-\pi}^{+\pi} \frac{P(r\,e^{i\,\varphi})}{r^{N}\,e^{i\,N\,\varphi}}\,d\varphi.
\end{equation}
Since $r\neq 0$ the maximum of the function under an integration in (\ref{e18}) is expected at $\varphi = 0$ taking into account the number of particles $N$ in the exponential function in (\ref{e17}). Hence, the asymptotic calculation of (\ref{e17}) requires to know the behavior of the function under the integration in (\ref{e18}) at $\varphi = 0$.  
Taking into account the minimum condition ({\ref{e12}) we have
$$\frac{\partial}{\partial\varphi}\left [ \frac{P(r\,e^{i\,\varphi})}{r^{N}\,e^{i\,N\,\varphi}}
\right ]_{\vert_{\varphi = 0}} = 0 .$$

The asymptotic form of the partition function is
 $$\ln Z_{N}\simeq -\sum_{f}\,\ln \left [1-\bar\mu_{0}\,e^{-F(f)\beta}\right ] - N\,\ln\bar\mu_{0} - \ln 2\sqrt {\kappa\,\pi\, N},$$
where $\kappa = - \Phi^{\prime} (\bar\mu_{0}) - \Phi^{\prime\prime} (\bar\mu_{0})$. Considering Eq. (\ref{e12}) at $\bar\mu = \bar\mu_{0}$, one can easily find 
\begin{equation}
 \label{e20}
\sum_{f} \bar n_{f} = \sum_{f} \frac{1}{\bar\mu_{0}^{-1}\, e^{F(f)\beta} -1} = N,
\end{equation}
where $\bar\mu_{0}$ can be defined from. Using (\ref{e20}) one can calculate the sum of quantum states up to singular point (CP) defined by the relation between $\bar\mu_{0}$ and $F\beta$, and large $N$.
We assume the large number $N$ in (\ref{e20}) which is correct if the dilatons are light. This is important in the sense of the proposal to condensed dark matter bosons in the early stage after (heavy ion) collisions. The latter in some sense corresponds to  Bose star formation as the lamps of BEC bounded by self-gravity [9].

{\bf 4. Critical temperature}

Consider the nonrelativistic model where the glueballs are produced in the volume $\Omega$ as a cube with the side of the length $L = \Omega^{1/3}$. The wave function of the glueball is 
$ \phi_{p} (q) = \Omega^{-1/2}\,e^{i\,q\,p}$, where $p^{\alpha} = (2\,\pi/L)\,n^{\alpha}$, $\alpha = 1,2,3$; $n^{\alpha} = 0, \pm 1, \pm 2, ...$ and the energy is $E_{p} = {\vert p\vert}^{2}/(2\,m_{\chi})$ (in the units with the Planck constant $h =1$).
In the limit $\Omega\rightarrow\infty$ and for $v = const$ we consider two cases:
high temperature case A), where $\bar\mu_{0}\,  e^{\mu\,Q\beta} < 1$, and
low $T$ case B), where $\bar\mu_{0}\, e^{\mu\,Q\beta} \sim 1$.

In case A) the function $\bar n_{f}$  is regular on $f$, and the sum $\sum_{f} \bar n_{f}$ is replaced by the integral, where the spectrum of $f$ is continuous at $\Omega\rightarrow\infty$:
$$\frac{1}{v} = \frac{1}{\Omega}\,\sum_{f}\bar n_{f}\rightarrow \frac{1}{(2\,\pi)^{3}}\int\bar n(f)\,d^{3} f. $$ Using the form of $\bar n(f)$ (\ref{e20}) we arrive at the equality:
\begin{equation}
 \label{e21}
 \int_{0}^{\infty}\frac{x^{2}\,dx}{\bar\mu_{0}^{-1}\,e^{-\mu\,Q\beta}\,e^{x^{2}} -1} = \frac{2\,\pi^{2}}{v}\,{\left (\frac{\beta}{2\,m_{\chi}}\right )}^{3/2}. 
\end{equation}
The integral in l.h.s. of (\ref{e21}) increases if $\mu\,Q + T \ln\bar\mu_{0} \rightarrow 0$ that will allow one to find an inequality
$$\frac{2\,\pi^{2}}{v}\,{\left (\frac{\beta}{2\,m_{\chi}}\right )}^{3/2}  < \int_{0}^{\infty}\frac{x^{2}\,dx}{e^{x^{2}} -1} . $$
The case A) is realised  when  the temperature $T$ exceeds the critical one, $T >T_{c}$, where 
$$T_{c} = \frac{1}{2\,m_{\chi}} {\left (\frac{2\,\pi^{2}}{v\,B}\right )}^{2/3}, \,\,\, B = \frac{\sqrt{\pi}}{4}\cdot 2,612... , \,\,\,\,m_{\chi} \neq 0.$$
One can easily find the correlation length $\xi$  is defined by $\mu$, has the dependence of $N$ (through $v$), and the singular behavior of $\xi$ is governed by the ground state $\bar\mu_{0}$:
\begin{equation}
 \label{e199}
\xi = 2\mu\,Q \left (\frac{v\,B}{2\,\pi^{2}}\right )^{2/3}\,\ln^{-1}\left (\frac{1}{\bar\mu_{0}}\right ). 
\end{equation}
Actually, $\xi\rightarrow\infty$ at $\bar\mu_{0}\rightarrow 1$ that means CP ($\bar\mu_{0} = \bar\mu =1$).


In case B) our interest is in small $\vert p\vert \leq\delta$ (maximal $N$), where CP is approached. Here
$$\frac{1}{\Omega}\sum_{\vert p\vert \leq\delta} \bar n_{p} = \frac{1}{v} - \frac{1}{\Omega}\sum_{\vert p\vert \geq\delta} \bar n_{p}, $$
where 
$$\lim_{\delta\rightarrow 0,\,\,N\rightarrow\infty} \frac{1}{\Omega}\sum_{\vert p\vert \leq\delta} \bar n_{p} = \frac{1}{v} - \frac{1}{(2\,\pi)^{3}}\int_{\vert p\vert \geq\delta} \frac{d^{3} p}{e^{E_{p}\beta} -1} = 
\frac{1}{v} \left [1-{\left (\frac{\beta_{c}}{\beta}\right  )}^{3/2} \right ]. $$
Hence, the case B) takes place for $T < T_{c}$, where the only part of total number of particles proportional $\sim (\beta_{c}/\beta)^{3/2}$ is distributed on all the spectrum of momenta. The rest one $\sim [ 1- (\beta_{c}/\beta)^{3/2}]$ is the scalar condensate.

Now, one can connect the results for the fluctuations of $\chi$ to the fluctuations of observable quantities. For this, we suppose the particles are in the local volume $V$ which is (much) less than $\Omega$. The number of particles $n_{V}$ in $V$ is $\sum_{1\leq j\leq N} \hat n_{V} (q_{j})$, where $\hat n_{V} (q) = 1$ if $ q\in V$, and $\hat n_{V}( q) = 0$ otherwise.  The volume $V$ is defined by the geometry of the experiment. 
The event-by-event fluctuation of particle density is a  function of $\xi$
\begin{equation}
 \label{e27}
\langle {(n_{V} - \langle n_{V}\rangle )}^{2}\rangle = \langle n_{V}\rangle 
\left [ 1 + \frac{\sqrt {2}\,v}{\pi^{2}}\, \left (\frac{T}{\xi}\right )^{3/2}\int_{0}^{\infty} \frac{ x^{2}\,dx}{(\bar\mu_{0}^{-1}\,e^{-\mu\,Q\,\beta}\,e^{x^{2}} -1)^{2}} \right ], 
\end{equation}
where $\langle n_{V} \rangle = V/\Omega$. The non-monotonous behaviour of particle density fluctuation is evident 
at CP and it is independent on $\xi$
$$\langle {(n_{V} - \langle n_{V}\rangle )}^{2}\rangle = \langle n_{V}\rangle 
\left [ 1 + B^{-1}\int_{0}^{\infty} \frac{ x^{2}\,dx}{(e^{x^{2}} -1)^{2}} \right ]. $$

{\bf 5. Primary photons at CP }


In the exact scale symmetry, $\chi$ couples to SM particles through the trace of $\theta_{\mu\nu}$
\begin{equation}
\label{e29}
 L = \frac{\chi}{f_{\chi}} \left ( \theta^{\mu}_{\mu_{tree}} +  \theta^{\mu}_{\mu_{anom}}\right ),
\end{equation}
where the first term in (\ref{e29}) is (contributions from heavy quarks and heavy gauge bosons are neglected)
$$\theta^{\mu}_{\mu_{tree}} = -\sum_{q} [m_{q} + \gamma_{m}(g) ]\bar q q  - \frac{1}{2} m_{\chi}^{2}\chi^{2} +\partial_{\mu}\chi\partial^{\mu}\chi, $$
 $q$ is a quark field  with the mass $m_{q}$; $\gamma_{m}$ are the corresponding anomalous dimensions. In contrast to SM, the dilaton couples to massless gauge bosons even before running any SM particles in the loop, through the trace anomaly.
The latter has the following term in (\ref{e29})  for photons and gluons:
$$\theta^{\mu}_{\mu_{anom}} = -\frac{\alpha}{8\,\pi}\, b_{EM}\, F_{\mu\nu}F^{\mu\nu} - 
\frac{\alpha_{s}}{8\,\pi}\sum_{i}\, b_{0_{i}}\, G_{\mu\nu}^{a}G^{{\mu\nu\,a}},$$
where $\alpha$ is the fine coupling constant, $b_{EM}$ and $b_{0_{i}}$ are the coefficients of electromagnetic (EM) and QCD $\beta$ functions, respectively. If the strong (and EM) interactions are embedded in the conformal sector the following relation for light and heavy particles sectors is established above the scale $\Lambda$ (in UV): 
$\sum_{light} b_{0} = - \sum_{heavy} b_{0}$, where the mass of $\chi$ splits the light and heavy states. The anomaly (non-perturbative) term for gluons in (\ref{e29})
$$\frac{\alpha_{s}}{8\,\pi}\,b^{light}_{0}  G_{\mu\nu}^{a}G^{{\mu\nu\,a}} = \frac{\beta (g)}{2\,g} G_{\mu\nu}^{a}G^{{\mu\nu\,a}}, \,\,\, b^{light}_{0} = -11 +\frac{2}{3}n_{L} $$
is evident, where the only $n_{L}$ particles lighter than $\chi$ are included in  $\beta$ - function, $\beta (g) = b^{light}_{0}\,g^{3}/(16\, \pi^{2})$, $ g^{2} = 4\pi\alpha_{s}$. For $m_{\chi}\sim O(\Lambda)$ one has $n_{L} = 3$ that indicates about 14 times increase of the dilaton-gluon-gluon coupling strength compared to that of the SM Higgs boson.

The light dilaton operates with low invariant masses, where two photons are induced effectively by gluon operators. In the low-energy effective theory, valid below the conformal scale $\Lambda_{conf} = 4 \pi f_{\chi}$, at small transfer-momentum $q$,  
$\langle \gamma\gamma\vert \theta^{\mu}_{\mu} (q)\vert 0\rangle\simeq 0$  [10]
and 
$$\langle \gamma\gamma\vert \frac{b^{light}_{0}\,\alpha_{s}}
{8\,\pi} G^{a}_{\mu\nu} G^{\mu\nu\,a}\vert 0\rangle = - \langle \gamma\gamma\vert \frac{b_{EM}\,\alpha}{8\,\pi} F_{\mu\nu} F^{\mu\nu}\vert 0\rangle, \,\, \vec q = 0.$$
The partial decay width $\chi\rightarrow\gamma\gamma$ is
$$ \Gamma (\chi\rightarrow\gamma\gamma)\simeq \left (\frac{\alpha\,F_{anom}}{4\,\pi}\right )^{2}\,\frac{m_{\chi}^{3}}{16\,\pi\,f_{\chi}^{2}},$$
where the only CA does contribute through 
$$F_{anom} = -(2\,n_{L}/3) (b_{EM}/b^{light}_{0}), \,\,\,b_{EM} = -4\sum_{q:u,d,s} e^{2}_{q} = -8/3, $$ 
$e_{q}$ is the charge of the light quark. In the vicinity of IRFP there are fluctuations of dilaton field with the mass which is given by $m_{\chi}\simeq \sqrt{1-N_{f}/N^{c}_{f}}\Lambda$ [3], where $N^{c}_{f}$ is the critical value of $N_{f}$ corresponding to $\alpha^{c}_{s}$ at which the chiral symmetry is breaking and the confinement is emerged. In order to estimate $\Gamma(\chi\rightarrow\gamma\gamma)$ we take $f_{\chi}\simeq\Lambda$. 
When one approaches the CP the absolute value of $F_{anom}$ decreases due to increasing on $b^{light}_{0}$ as $n_{L}\rightarrow 0$. The second-order phase transition is characterised by the limits $N_{f}\rightarrow N^{c}_{f}$ and $\Lambda\rightarrow 0$, hence no primary photons should be evident through a detector. 
In the IR ($\alpha^{\star}_{s} > \alpha^{c}_{s}$) one can estimate the fluctuation rate (as an observable) relevant to  primary photons with $N_{f}$ and $n_{L}$ 
\begin{equation}
\label{e30}
r_{\chi} = C_{EM}\,\Gamma(\pi^{0}\rightarrow\gamma\gamma)\,\left (\frac{\Lambda}{n_{L}}\right )^{2}\,\xi^{3}, 
\end{equation}
where at large distances we use the effective d.o.f. in terms of neutral $\pi^{0}$ - mesons, and $\Gamma(\pi^{0}\rightarrow\gamma\gamma)$ is the partial decay width $\pi^{0}\rightarrow\gamma\gamma$; $ C_{EM} = (4\,\pi)^{3\,}[3\,b_{0}^{light}/(\alpha\, b_{EM})]^{2}$; $\xi$ is given by (\ref{e199}).
Actually, $r_{\chi}$ (\ref{e30}) in terms of confinement scale $\Lambda$ is scheme independent. 
The result (\ref{e30}) is consistent with the physical pattern where dilaton is emerged at scales $\geq \Lambda$ as well as $\pi^{0}$'s and other light quark bound states.
It is easily to find that at CP $r_{\chi}\rightarrow\infty$ when the number of light quarks $n_{L}\rightarrow 0$ as well as the fluctuation length $\xi$ is sharply increasing. The latter is the consequence of very small mass $m_{\chi}$ of the dilaton at $N_{f}\rightarrow N_{f}^{c}$
as well as  $\alpha^{\star}_{s}\rightarrow \alpha^{c}_{s}$.
 Thus, one can expect to find the non-monotonous raising on fluctuations of primary photons once  is going away from  UV to IR. The measurement of photon fluctuations can be used to determine whether the quantum system is in the vicinity of CP or not.

{\bf 6. Conclusion}

To conclude, the novel approach to an approximate scale symmetry breaking up to the phase transition at the critical point is suggested. The possible determination of the phase boundary between the confinement-deconfinement border and the high $T$ plasma phase can be seen inside the conformal window.

We find the CP is achieved at higher $\mu$ (case B) with smaller particle momentum (and, hence, the energy). In the vicinity of CP one has the scalar condensate with the sharp increasing of particle density at CP (\ref{e27}). 

The CP can be found as those followed by IRFP where the primary photons are detected. The origin of these photons is CA through the decays of the dilatons. When the incident energy scans from high to low values, one can find the non-monotonous behaviour in fluctuations of  primary photons: 
these fluctuations grow in the IR to become large at CP (\ref{e30}). 
An experimental information about the location of CP for the given experimental conditions is obtained by measuring the ratios of $\gamma$-quanta yields and compared (fitting) to known model with $T$ and $\mu$.

\end{document}